\documentclass{openjournal}
\usepackage{graphicx}
\usepackage{amsmath}
\usepackage{amssymb}
\usepackage{xspace}
\usepackage[hidelinks]{hyperref}
\usepackage{hyperref}
\usepackage{orcidlink}

\def \MIDAS {Michigan Institute for Data and AI in Society, University of Michigan, Ann Arbor, MI 48109, USA}
\def \Physics {Department of Physics, University of Michigan, Ann Arbor, MI 48109, USA}
\def \CfA {Center for Astrophysics | Harvard \& Smithsonian, 60 Garden Street, Cambridge, MA 02138, USA}

\usepackage{xcolor}
\definecolor{Maize}{HTML}{FFCB05}
\definecolor{Blue}{HTML}{00274C}
\definecolor{TappanRed}{HTML}{9A3324}
\definecolor{RossOrange}{HTML}{D86018}
\definecolor{RackhamGreen}{HTML}{75988d}
\definecolor{WaveFieldGreen}{HTML}{A5A508}
\definecolor{TaubmanTeal}{HTML}{00B2A9}
\definecolor{ArboretumBlue}{HTML}{2F65A7}
\definecolor{A2Amethyst}{HTML}{702082}
\definecolor{MatthaeiViolet}{HTML}{575294}
\definecolor{UMMATan}{HTML}{CFC096}
\definecolor{BurtonTowerBeige}{HTML}{9B9A6D}
\definecolor{AngelHallAsh}{HTML}{989C97}
\definecolor{LawQuadStone}{HTML}{655A52}
\definecolor{PumaBlack}{HTML}{131516}
\hypersetup{
    colorlinks,
    linkcolor={red!50!black},
    citecolor={blue!50!black},
    urlcolor=ArboretumBlue
}

\begin{document}

\title{Improved Initial Guesses for Numerical Solutions of Kepler's Equation}

\shorttitle{Improved Initial Guesses for Numerical Solutions of Kepler's Equation}
\shortauthors{Napier}

\author{\vspace{-1.2cm}Kevin~J.~Napier\,\orcidlink{0000-0003-4827-5049}}
\affiliation{\MIDAS}
\affiliation{\Physics}
\affiliation{\CfA}

\begin{abstract}
Numerical solutions of Kepler's Equation are critical components of celestial mechanics software, and are often computation hot spots. This work uses symbolic regression and a genetic learning algorithm to find new initial guesses for iterative Kepler solvers for both elliptical and hyperbolic orbits. The new initial guesses are simple to implement, and result in modest speed improvements for elliptical orbits, and major speed improvements for hyperbolic orbits.
\end{abstract}


\section{Introduction}
\label{sec:intro}

Kepler's Equation, which describes the relation between an elliptical orbit's mean anomaly ($M$) and its eccentric anomaly ($E$), is given by
\begin{equation}
    M = E - e\sin{E}
    \label{eq:kepler}
\end{equation}
where $e$ is the orbit's eccentricity. Equation (\ref{eq:kepler}) is transcendental, meaning that it is impossible to solve for $E$ in terms of elementary functions. The lack of such a solution is unfortunate, as Kepler's equation is one of the most important equations in celestial mechanics, enabling the analytic propagation of bodies along Keplerian orbits.

\cite{philcox2021} found a solution to Equation (\ref{eq:kepler}) in the form of a ratio of contour integrals. In practice, however, evaluating the two contour integrals is slower than solving the equation using numerical root-finding algorithms such as Newton's method. Furthermore, iterative numerical root finding schemes have the advantage that they can achieve solutions with arbitrarily high precision by simply continuing to iterate. The contour integral solution, on the other hand, requires the integral to be re-calculated at higher resolution in order to reach higher precision. So despite the existence of an analytic solution, iterative root-finding solutions are still preferable. 

Given the high frequency with which the equation must be solved in software, it is always worth considering whether numerical solutions can be improved. The speed at which a numerical root-finding algorithm converges on a solution depends on the quality of the initial guess, with a trade-off between more complicated initial guesses and extra iterations. Toward that end, this paper revisits the problem of finding an initial guess for both the elliptical and hyperbolic forms of Kepler's Equation. Section \ref{sec:methods-and-results} motivates and implements a machine learning approach to the elliptical form of the problem, and Section \ref{sec:benchmarks} explores the effectiveness of the solution. Section \ref{sec:hyperbolic} implements the same approach for hyperbolic orbits. The paper concludes with a brief discussion in Section \ref{sec:conclusions}.


\section{Methods and Results}
\label{sec:methods-and-results}

There are many ways to go about making initial guesses for solutions of Kepler's Equation. The traditional approach has been to make an analytic estimate using a series expansion such as $E = M + e \sin{M}$, or an ad-hoc expression such as $E = M + 0.85 e$ \citep{danby}. \citet{markley1995} found an analytic expression that has a relative error smaller than $10^{-18}$ everywhere in the solution space. However, the evaluation of the expression usually takes longer than using an iterative root finding algorithm with a simple initial guess.

Alternatively, it is possible to use a data-driven approach to make an initial guess. For example, one could pre-calculate a fixed grid of solutions, and then interpolate the grid to find initial guesses at arbitrary locations in the parameter space. Similarly, one could train a neural net to closely approximate the function. Both approaches have a major drawback in that they require some nonzero amount of data to be loaded into memory. In a perfect contrived experiment where the required data could all be loaded into the CPU cache, and be guaranteed to remain there, such approaches would likely perform well. In practice, however, the data would be passed around between the CPU cache, the stack, and the heap, leading to slow memory access times and cache misses, thus degrading the overall performance of the algorithm. Because of these practical limitations, analytic expressions for initial guesses are likely to yield better performance.

By using modern machine learning techniques, it is possible to simultaneously leverage the power of analytic and data-driven approaches. A combination of symbolic regression with a genetic learning algorithm is particularly well-suited to the problem at hand, as it enables an efficient exploration of the space of all possible functions to find one that offers a close approximation of the solution to Equation (\ref{eq:kepler}), and yields an analytic expression rather than a black box solution (such as a neural net). The basic idea of this approach is to use a set of pre-defined operations to construct some function, $f$, that minimizes the residuals of $E = f(e, M)$ over the full parameter space. In the interest of computation speed, I limited the available mathematical operations to addition, subtraction, multiplication, \texttt{min}, and \texttt{max}, and discouraged the algorithm from searching for long expressions by using a parsimony penalty. I allowed the algorithm to use the variables $e$, $M$, and $\sin{M}$, as well as scalar quantities of order unity. It is also useful to restrict the parameter space to $e \in [0, 1)$ $\times$ $M \in [0, \pi]$, by exploiting the symmetry in $M$.\footnote{The fits in this paper all used a uniform sampling of points in the parameter space, but it is possible to weight the fit by using some nonuniform sampling if one is particularly interested in fast solutions for certain classes of orbits. For example, I tried sampling points logarithmically in $e$, as well as restricting the parameter space to $e \in [0, 0.2]$ to favor lower eccentricity orbits, though I was unable to find an improvement in either case.}

For a high parsimony penalty (i.e., forcing the algorithm to yield terse expressions), the algorithm converged to initial guesses $E = M + e \sin{M}$ and $E = M + 0.71 e$, which are consistent with the canonical initial guesses in the literature \citep{danby}. By allowing for slightly more complicated expressions, the algorithm converged on the initial guess
\begin{equation}
    E = e \sin M + \texttt{max}\left\{M, e (\sin M + 0.591)\right\}.
    \label{eq:new-guess}
\end{equation}
It is interesting to note that this new initial guess is identical to the canonical initial guess of $E = M + e \sin{M}$ when $M \geq e (\sin M + 0.591)$. It seems that the algorithm used the \texttt{max} function to divide the parameter space into two regions, thus making each region easier to fit.\footnote{It is possible in principle to manually emulate this behavior by dividing the parameter space into different regions, each with its own initial guess. However, such an implementation would lead to branching code, causing degraded performance. The \texttt{max} operation used in this solution does not have the same drawback.}

\section{Benchmarks}
\label{sec:benchmarks}

\begin{figure}[ht]
    \centering
    \includegraphics[width=0.95\textwidth]{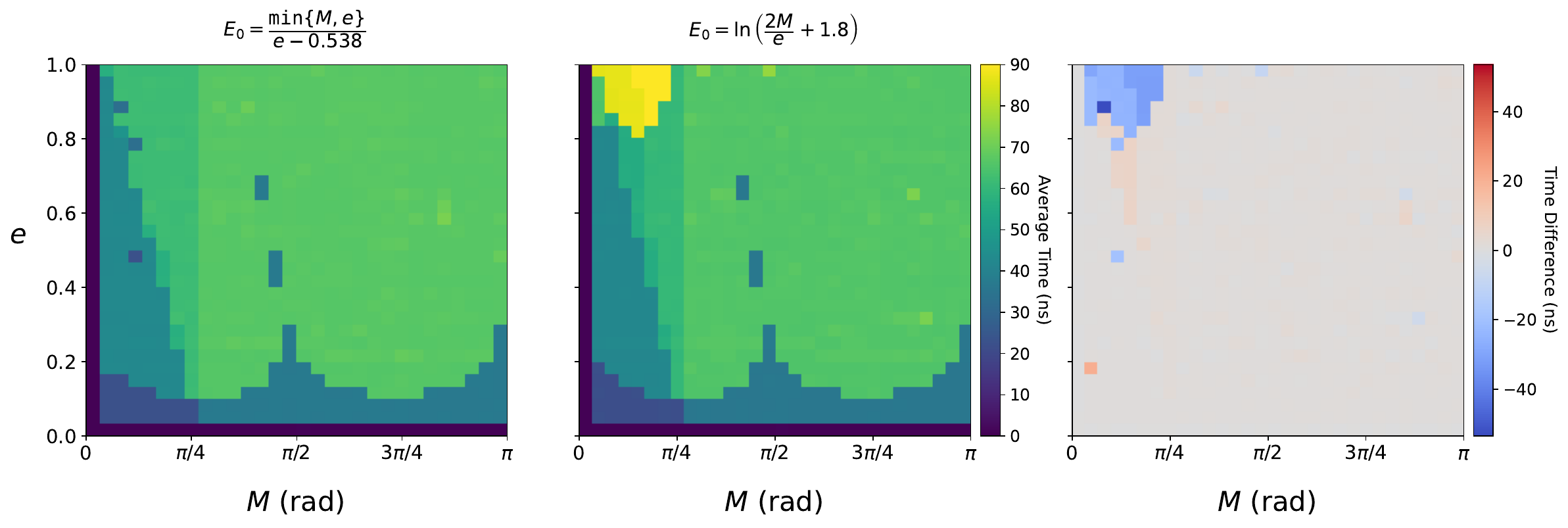}
    \includegraphics[width=0.95\textwidth]{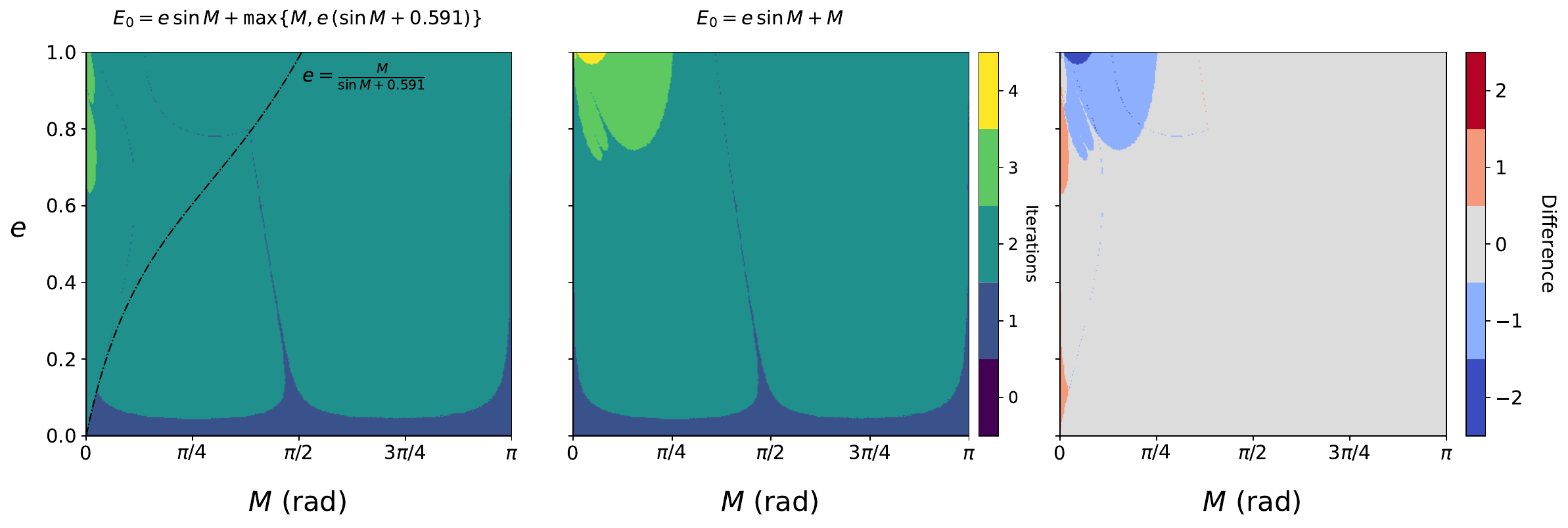}
    \caption{(Top) Average time taken to converge on a solution to Kepler's Equation at a precision of $10^{-15}$, parameterized as a function of $e$ and $M$. The new initial guess (left) results in faster convergence than the canonical initial guess (right) for high-eccentricity orbits near pericenter. The performance is similar in the rest of the parameter space. (Bottom) Number of iterations taken to converge on a solution to Kepler's Equation at a precision of $10^{-15}$.}
    \label{fig:benchmark}
\end{figure}

Equation (\ref{eq:new-guess}) represents a low-cost expression for an initial guess to the solution of Kepler's Equation that is more accurate than traditional expressions. In practice, however, the efficacy of the initial guess matters less than the actual speed at which the iterative algorithm converges on a solution, so it is still yet to be seen whether this new initial guess yields any benefit. Figure \ref{fig:benchmark} shows empirical benchmarks to compare the effectiveness of Equation (\ref{eq:new-guess}) against that of the canonical guess, $E = M + e \sin{M}$, using a quartic root-finding algorithm.\footnote{The time benchmarks in Figure \ref{fig:benchmark} were all calculated robustly using the \texttt{criterion} library in \texttt{Rust} \citep{criterion.rs}.} It is clear that Equation (\ref{eq:new-guess}) performs identically to the canonical guess over most of the parameter space, but does offer a speed improvement for highly eccentric orbits close to pericenter.

\section{Hyperbolic Orbits}
\label{sec:hyperbolic}

\begin{figure}[ht]
    \centering
    \includegraphics[width=0.95\textwidth]{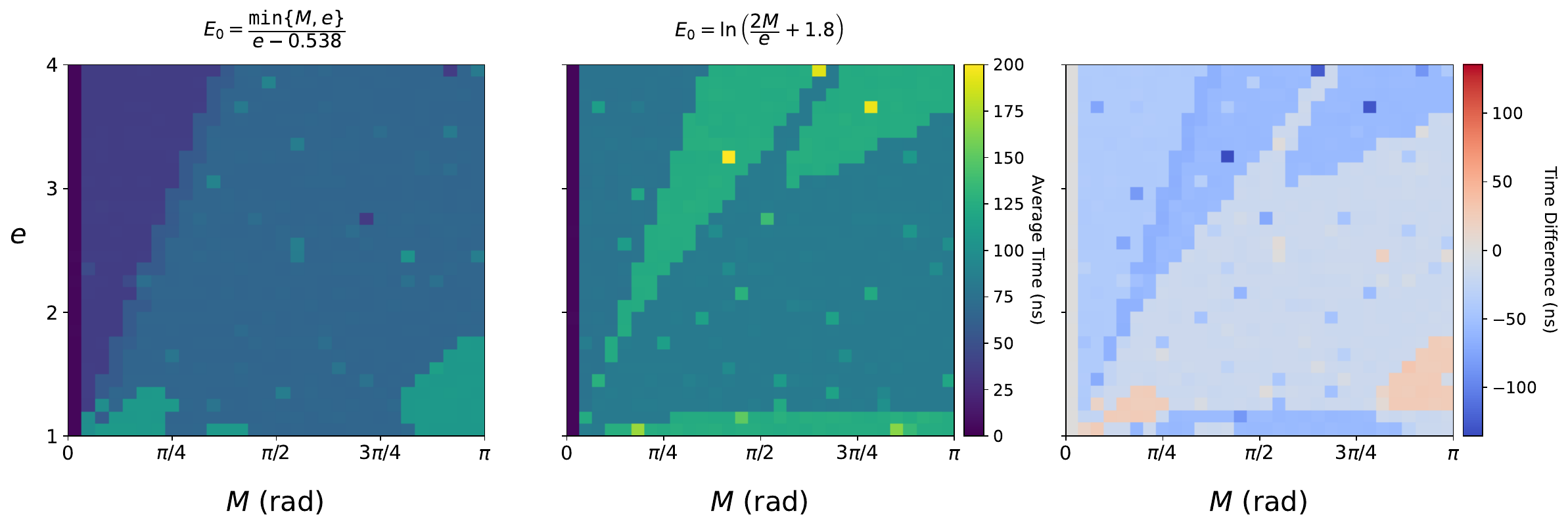}
    \includegraphics[width=0.95\textwidth]{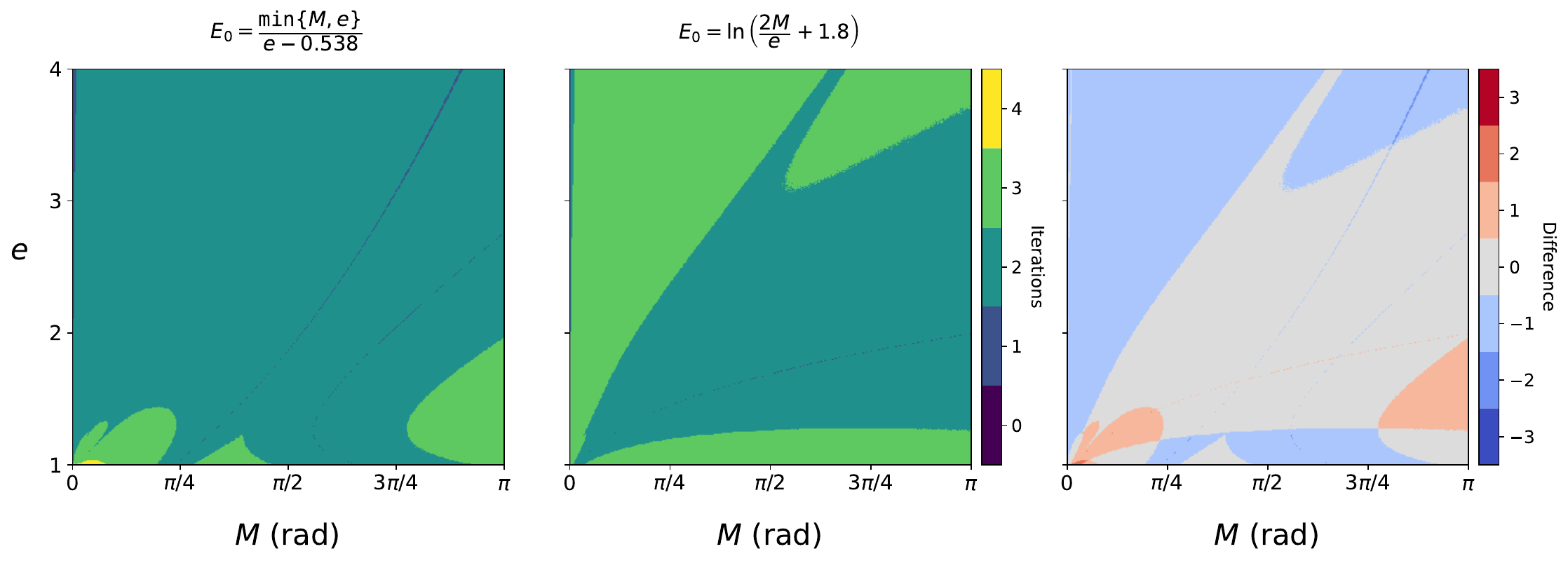}
    \caption{(Top) Average time taken to converge on a solution to Kepler's Equation at a precision of $10^{-14}$, parameterized as a function of $e$ and $M$. There is some noise in these benchmarks, but it is clear that the new initial guess (left) results in faster convergence than the canonical initial guess (right) for most of the parameter space. (Bottom) Number of iterations taken to converge on a solution to Kepler's Equation at a precision of $10^{-14}$.}
    \label{fig:hyperbolic-benchmark}
\end{figure}

For hyperbolic orbits, Kepler's Equation is given by 
\begin{equation}
    M = e \sinh{E} - E
\end{equation}
where $E$ is now referred to as the hyperbolic anomaly. The usual initial guess for hyperbolic orbits is given by $E = \ln(2M/e + 1.8)$ \citep{danby}. Using the variables $e$ and $M$, scalars of order unity, and the operations of addition, subtraction, multiplication, division, \texttt{min}, and \texttt{max}, the previously-described algorithm converges on the very simple expression
\begin{equation}
    E = \frac{\texttt{min}\left\{M, e \right\}}{e - 0.538}.
\end{equation}
It is clear from the benchmarks in Figure \ref{fig:hyperbolic-benchmark} that this new initial guess is an improvement over the canonical initial guess. It also appears that the improvement holds for orbits with arbitrarily large eccentricity.

\section{Conclusions}
\label{sec:conclusions}
This work used a combination of symbolic regression and a genetic learning algorithm to make a modest improvement to numerical solutions of the elliptical case of Kepler's Equation, and a more significant improvement to numerical solutions of the hyperbolic case. The use of this particular machine learning algorithm is particularly well-suited to scientific applications because the results are interpretable (i.e., not a black box), and it is easy to prevent the algorithm from overfitting.

It is possible that better initial guesses for Kepler's Equation exist; the purpose of this paper is not to be the final word on the subject, but rather to improve on the current state-of-the-art, and to introduce a technique for finding more complicated, yet still easily-calculable, initial guesses for iterative root-finding solvers. The functions I found in this work are simple; for more complex functions, it is possible for initial conditions to cause iterations to fail to converge, due either to floating point issues or oscillatory behavior.

Finally, I note that the use of $\sin M$ in the elliptic initial guess may make the results machine dependent, as the initial guess will depend on the implementation of $\sin$ in the chosen math library. In principle, it may be possible to avoid this issue by restricting the initial guess to only use additions and multiplications; since the trigonometric functions can be approximated as Taylor series, this approach may lead to improved performance. Toward this end, I tried restricting symbolic regressor to use only $e$, $M$, and the functions in the set $\{+,\times\}$, but I did not find a improved solution. I also tried providing $M^2 / 2$ and $M^3 / 6$ to the regressor in order to coax out the form of the Taylor expansion, but still did not find an improved solution. Of course it is possible to manually override the $\sin$ function with a series approximation, which may provide an easy speedup. 

\section*{Acknowledgements}
This project is supported by Schmidt Sciences, LLC. I thank Hanno Rein, Fred Adams, and Hsing-Wen Lin for feedback that has improved this study.

\bibliographystyle{aasjournal}
\bibliography{references}

\end{document}